\documentclass[a4paper,onecolumn,11pt,accepted=2018-05-04]{quantumarticle}
\usepackage[utf8]{inputenc}
\pdfoutput=1
\usepackage[T1]{fontenc}
\usepackage[UKenglish]{babel}
\usepackage{hyperref}
\usepackage{graphicx}
\usepackage{amsmath,amssymb,amsthm}
\usepackage{xspace}
\usepackage{mathtools}
\usepackage{dsfont}
\usepackage{xcolor}
\usepackage{paralist}
\usepackage{array}
\usepackage{lipsum}
\usepackage{subfig}

\usepackage{cite}
 
\DeclareMathOperator{\tr}{Tr}

\newcommand{\id}{\mathds{1}}

\newcommand{\bra}[1]{\left\langle #1 \right|}
\newcommand{\ket}[1]{\left| #1 \right\rangle}

\newcommand{\Proj}[1]{[[#1]]}

\newtheorem*{theorem*}{Theorem}

\newtheorem{assumption}{Assumption}

\newcommand{\lin}{{\cal L}}

\usepackage[normalem]{ulem} 
\usepackage{cancel} 

\begin{document}
 
\title{Causation does not explain contextuality}

\author{Sally Shrapnel}
\email{s.shrapnel@uq.edu.au}
\orcid{0000-0001-8407-7176}
\affiliation{Centre for Engineered Quantum Systems, School of Mathematics and Physics, The University of Queensland, St Lucia, QLD 4072, Australia}
\author{Fabio Costa}
\email{f.costa@uq.edu.au}
\orcid{0000-0002-6547-6005}
\affiliation{Centre for Engineered Quantum Systems, School of Mathematics and Physics, The University of Queensland, St Lucia, QLD 4072, Australia}
\date{\today}
\begin{abstract}

Realist interpretations of quantum mechanics presuppose the existence of elements of reality that are independent of the actions used to reveal them. Such a view is challenged by several no-go theorems that show quantum correlations cannot be explained by non-contextual ontological models, where physical properties are assumed to exist prior to and independently of the act of measurement. However, all such contextuality proofs assume a traditional notion of causal structure, where causal influence flows from past to future according to ordinary dynamical laws. This leaves open the question of whether the apparent contextuality of quantum mechanics is simply the signature of some exotic causal structure, where the future might affect the past or distant systems might get correlated due to non-local constraints. Here we show that quantum predictions require a deeper form of contextuality: even allowing for arbitrary causal structure, no model can explain quantum correlations from non-contextual ontological properties of the world, be they initial states, dynamical laws, or global constraints.

\end{abstract}
\maketitle

\section*{Introduction} The appeal of an operational physical theory is that it makes as few unwarranted assumptions about nature as possible. One simply assigns probabilities to experimental outcomes, conditioned on the list of experimental procedures required to realise these outcomes. Ideally, such operational theories are \emph{minimal}: procedures that cannot be statistically discriminated are given the same representation in the theory. Quantum mechanics is an example of such a minimal operational theory: all the statistically significant information about the preparation procedure is contained in the quantum state, and the probability of an event (labelled by a Positive Operator Valued Measure (POVM) element) does not depend on any other information regarding the manner in which the measurement was achieved (such as the full POVM). However, one of the most debated questions in the foundations of the theory is whether one can go beyond this statistical level and also provide an \emph{ontological} description of some actual state of affairs that occurs during each run of an experiment. That is, a statement about the world that tells us what is responsible for the observed experimental outcomes. 

The task of providing such an ontological model for quantum theory has proven to be exceedingly difficult. A plethora of no-go theorems exists that describe the various natural assumptions one must forgo in order to produce an ontological model that accords with experiment. One such caveat is \emph{non-contextuality}. Ultimately an \emph{apriori} assumption, non-contextual theories posit the existence of physical properties that do not depend on the way they are measured. There is a large literature discussing the various ways one may wish to cash out this notion more precisely. Broadly speaking, non-contextuality no-go theorems fall into two distinct categories. Kochen-Specker style proofs show that quantum measurements cannot be regarded as deterministically uncovering pre-existing, or ontic, properties of systems~\cite{kochen67, bell66, cabello08}. Spekkens style proofs, on the other hand,  show that one cannot explain quantum statistics via ontological properties that  mirror the context-independence seen at the operational level~\cite{spekkens05, Montina2011, kunjwal2016, Mazurek2016, Schmid2017}. While both approaches are well justified and have led to interesting and relevant results, our own definition of non-contextuality is more closely related to the latter. This particular view of non-contextuality can more broadly be seen as an analogue of the no fine-tuning argument from causal modelling~\cite{cavalcanti2017}, an analogue of Leibniz's principle of the Identity of Indiscernibles~\cite{spekkens05, kunjwal2016}, and a methodological assumption akin to Occam's razor.

Non-contextuality no-go theorems are not merely of foundational interest but can also serve as security proofs for a range of simple cryptographic scenarios~\cite{chailloux2016, spekkens09}, can herald a quantum advantage for computation~\cite{Howard2014}, and also for state discrimination~\cite{Schmid2017}. Such results, however, require the assumption of a fixed background causal structure; at the very minimum, a single causal arrow from preparation to measurement. This leaves open the question of whether one can produce a non-contextual ontological model by allowing for a suitably exotic causal structure.  Some authors attempt to explain quantum correlations by positing backwards-in-time causal influences~\cite{price2012, priceWharton2015, Evans01062013, Wharton2014, Aharonov2016, Leifer2016, Sutherland17}, while others claim it is the existence of non-local constraints that does the explanatory work~\cite{carati1999nonlocality, Weinstein2009}. The rationale in both cases is that non-contextuality could emerge naturally in such models: physical properties might well be ``real'' and ``counterfactually definite'', but depend on future or distant measurements because of some physically motivated---although radically novel---causal influence.  Such proposals do not fit neatly within the classical causal modelling framework, and so are not ruled out by recent work in this direction~\cite{wood2015, cavalcanti2017}, nor by any of the existing no-go theorems. 

In this paper, we characterise a new ontological models framework to prove  that even if one allows for \emph{arbitrary} causal structure, ontological models of quantum experiments are necessarily contextual. Crucially, what is contextual is not just the traditional notion of ``state'', but any supposedly objective feature of the theory, such as a dynamical law or boundary condition. Our finding suggests that \emph{any} model that posits unusual causal relations in the hope of saving ``reality'' will necessarily be contextual.  
Finally, this work also represents a possible approach to how we ought to think of the generalised quantum processes of recent work~\cite{gutoski06, chiribella08, Chiribella2008, chiribella09b, Bisio2011, Bisio2014, oreshkov12, modioperational2012, Leifer2013, Ringbauer2015, pollockcomplete2015, costa2016, Allen2016, Milz2016, shrapnel2017}. It is clear that any ontological reading of such processes will have to contend with the spectre of contextuality.

The paper is organised as follows. In section~\ref{OntModels} we present the traditional ontological models framework and clarify the rationale behind retrocausal explanations of quantum statistics. In section~\ref{opPrimitives} we introduce and justify the four primitive elements required to define our operational model: local regions, local controllables, outcomes and an environment. In section~\ref{opEquiv} we define the three classes of operationally indistinguishable elements: events, instruments and processes. In Section~\ref{ontModel} we characterise instrument and process non-contextuality according to these equivalence classes, and provide a generalised framework for a non-contextual ontological model. As this is the conceptual heart of our result, in Section~\ref{examples} we clarify the scope and applicability of this framework via three examples. Using standard quantum theory and results from previous work~\cite{oreshkov12, shrapnel2017}, in section~\ref{quantModel} we characterise an operational model that accords with the experimental predictions of quantum theory. Section~\ref{contradiction} puts these elements together to prove that one cannot produce an ontological model that is both process and instrument non-contextual and accords with the predictions of quantum theory. In Section~\ref{extension} we consider the constraints imposed on ontological models when one only assumes instrument non-contextuality. We finish with a discussion.

\section{An introduction to ontological models and retrocausal approaches.}\label{OntModels}

The ontological models framework assumes that systems possess well defined properties at all times~\cite{harrigan10,leifer2014quantum, Leifer2016}. The starting point is the very general claim that all experiments can be modelled operationally as sets of preparations, followed by transformations, followed by measurements, all performed upon some physical system. The set of all possible preparations, transformations and measurements is regarded as capturing the entire possibility space of any experiment and can be associated with the operational predictions of a particular theory. 
For example, an experiment can involve choices of possible preparation settings (labelled by the random variable $P$) and choices of possible measurement settings ($M$) with associated outcomes ($a$).\footnote{For this example we assume that any transformation between preparation and measurement is trivial.} An operational model then predicts probabilities for outcomes for all possible combinations of preparations and measurements: 
\begin{equation}
\forall a,M,P:~p(a|M,P).
\end{equation}

Such probabilistic predictions should coincide with the operational predictions of the theory in question. For example, in the case of quantum theory each preparation choice is modelled as a density operator ($\rho_P$) on a Hilbert space associated to a quantum system $(\mathcal{H}_A)$. Similarly, each measurement choice $M$ is associated with a positive operator valued measure $ \{E_{a|M} \}$, whose elements correspond to particular outcomes $a$.
The probabilities predicted by the theory are:
\begin{equation}
p(a |P, M) = \tr( \rho_P E_{a|M}).
\end{equation}

An \emph{ontological extension} of such an operational model further assumes that the system possesses well defined ontological properties between the time of preparation and measurement. Such properties are collectively known as the "ontic state" and typically denoted by $\lambda$. In the ontological models framework each preparation procedure $P$ is presumed to select a particular ontic state $\lambda$ according to a fixed probability distribution: $\mu_P(\lambda)$, and each measurement choice is presumed to output a particular outcome according to a fixed response function: $ \xi_{a|M}(\lambda)$. That is, (i) every preparation $P$ can be associated to a normalised probability distribution over the ontic state space $\mu_P(\lambda)$, such that $\int\mu_P (\lambda)d\lambda = 1$, and (ii) every measurement $M$, with outcomes $a$, can be associated to a set of response functions $\{\xi_{a|M}(\lambda)\}$ over the ontic states, satisfying $\sum_a \xi_{a|M}(\lambda) = 1$ for all $\lambda$.

As the ontic states are not directly observed, the operational statistics are obtained via marginalisation and we have:

\begin{equation}\label{OntMod}
\forall a,M,P:~p(a|M,P) = \int \xi_{a|M}(\lambda) \mu_P(\lambda) d \lambda,
\end{equation}
where for quantum theory:
\begin{equation}
\forall a,M,P:~\tr( \rho_P E_{a|M}) = \int \xi_{a|M}(\lambda) \mu_P(\lambda) d \lambda.
\end{equation}

The ontological models framework has been used in numerous works to clarify the manner in which quantum theory should be considered contextual~\cite{spekkens05, spekkens08, Montina2011, kunjwal2016, Mazurek2016, Schmid2017}. The key assumption is that one can infer ontological equivalence from operational equivalence: for example, if two preparation procedures produce the same distributions over outcomes for all possible measurements,  then any differences between them do not play a role in determining the ontic states of the system in question. Thus, the justification for why one can't distinguish between the two equivalent preparations at the \emph{operational} level is because there is no difference between the role the preparations play at the \emph{ontological} level. The view is that each use of a preparation device selects one from a set of possible ontic states according to exactly the same probability distribution in each run. Formally,  if $\forall M$ and outcome $a$ 
\begin{equation}
 p(a|M,P_1) = p(a|M,P_2),
\end{equation}
then both preparations specify the same distribution over ontic states:
\begin{equation}
\mu_{P_1}(\lambda) = \mu_{P_2} (\lambda).
\end{equation}
Similarly, if two measurements result in the same outcome statistics for all possible preparations then both measurements are represented by the same fixed response function. Formally,  if $\forall P$ and outcomes $a$ 
\begin{equation}
 p(a|M_1,P) = p(a|M_2,P),
\end{equation}
then both measurements specify the same distribution over ontic states:
\begin{equation}
\xi_{a|M_1}(\lambda) = \xi_{a|M_2}(\lambda).
\end{equation}
Thus, in a non-contextual ontological model one can account for operational statistics according to Eq.~\ref{OntMod}. Implicit in this model is the belief that the ontic state screens off the preparations from the measurements (a property also known as $\lambda$-mediation~\cite{Leifer2016}).
In~\cite{spekkens08} it was shown that one can use the assumptions above to derive a contextuality proof: no model of the form of Eq.~\ref{OntMod} can explain the statistics of quantum theory.

In this approach to contextuality~\cite{spekkens05, spekkens08, Montina2011, kunjwal2016, Mazurek2016, Schmid2017}, one assumes that ontic states determine correlations according to some fixed causal order.  Formally, this is captured by Eq.~\ref{OntMod}: the preparation is assumed to cause the selection of a particular ontic state $\lambda$ according to a fixed distribution $\mu(\lambda)$, and the measurement choice does not alter this value but merely determines the outcome probability, also according to a fixed probability distribution. This leaves open the question of whether one can explain the contextuality of quantum theory by postulating an alternative, retrocausal ontology: If the future can affect the past, then the state $\lambda$ could depend on the measurement setting $M$, and Eq.~\eqref{OntMod} would not be justified.

Generally speaking, retrocausal approaches posit the existence of backwards-in-time causal influences to explain quantum correlations. The stated appeal of such approaches is that the consequent explanations retain some element of our classical notion of reality: local causality, determinate ontology, and counterfactual definiteness. For example, 
Price and Wharton explain Bell correlations by including a  "zig-zag" of causal influence, passing via hidden variables that travel backwards in time from one measurement event to the source and then forwards in time to the distant measurement event~\cite{priceWharton2015}.
Although not explicitly stated, there is also one further assumption underlying these approaches: such causal influences follow some kind of law-like behaviour. That is, one would not expect the \emph{rules} by which such retrocausal influences propagate, or backward-in-time states evolve, to be completely ad hoc. 

As stated in the introduction, we follow the Spekkens-style approach and also define non-contextuality in terms of operational equivalences. Where we depart however, is in our particular choice of operational primitives. The usual primitives of preparations, transformations and measurements do not permit one to consider causal scenarios that move beyond the most simple causally ordered situations; in these models the notion of reality is defined in terms of properties that exist \emph{before} a measurement takes place. The underlying ontology is therefore assumed to follow some ordinary causal structure, akin to the directed acyclic graphs of causal models~\cite{Pearlbook}. In our model we wish to be able to consider more general situations, for example where we include \emph{any} possible global dynamics, causal structure, space-time geometry or global constraints. In order to provide this alternative perspective we consider the primitive operational elements to be sets of labelled local regions, locally controllable properties and an environment.

\section{Operational primitives}\label{opPrimitives}

We define an operational model of any experiment to consist of \emph{local labelled regions} ($A, B, C,\dots$) where one can perform \emph{controlled operations} that can be associated with \emph{outcomes}. The regions align with concepts such as local laboratories, communicating parties (e.g. Alice and Bob) and local space-time regions (similar, e.g., to the operational framework of~\cite{oreshkov15}). There is no \emph{apriori} assumption that these regions be "fixed" or preassigned in some manner; they are simply labels for the locus of a set of controlled operations. Controlled operations generalise the notion of preparations, measurements, transformations, and can include the addition or subtraction of ancillary systems. Examples include the orientation of a wave-plate, the instigation of a microwave pulse, and the use of a photo-detector. We call such local operations the \emph{local controllables}.  Each local controllable is represented as $\tilde{\mathcal{\mathfrak{I}}}^X$, where the superscript $X= A, B,\dots$ labels the associated region. We consider outcomes as labels associated to the result of choosing a particular local controllable; the outcomes for region $A$ are labelled $a=0,1,2,\dots$. Examples include the number of detected photons, the result of a spin measurement or the time of arrival of a photon. We allow the outcomes to have infinite possible values as this enables us to use the same variable for local controllables that have different numbers of possible outcomes.  In general however, we expect that only a finite number of such outcomes is associated with non-zero probability.

Finally, we consider all the possible properties that could account for correlations between outcomes in the local regions. These include any global properties, initial states, connecting mechanisms, causal influence, or global dynamics. We call this the \emph{environment}, $\tilde{W}$. Note that in our operational model environments and local controllables are by construction always uncorrelated. That is, if we see a property change in relation to a choice of local controllable we label this as an \emph{outcome} and do not classify it as part of the environment. 
\begin{figure}[ht]%
\centering
\includegraphics[width=0.7\columnwidth]{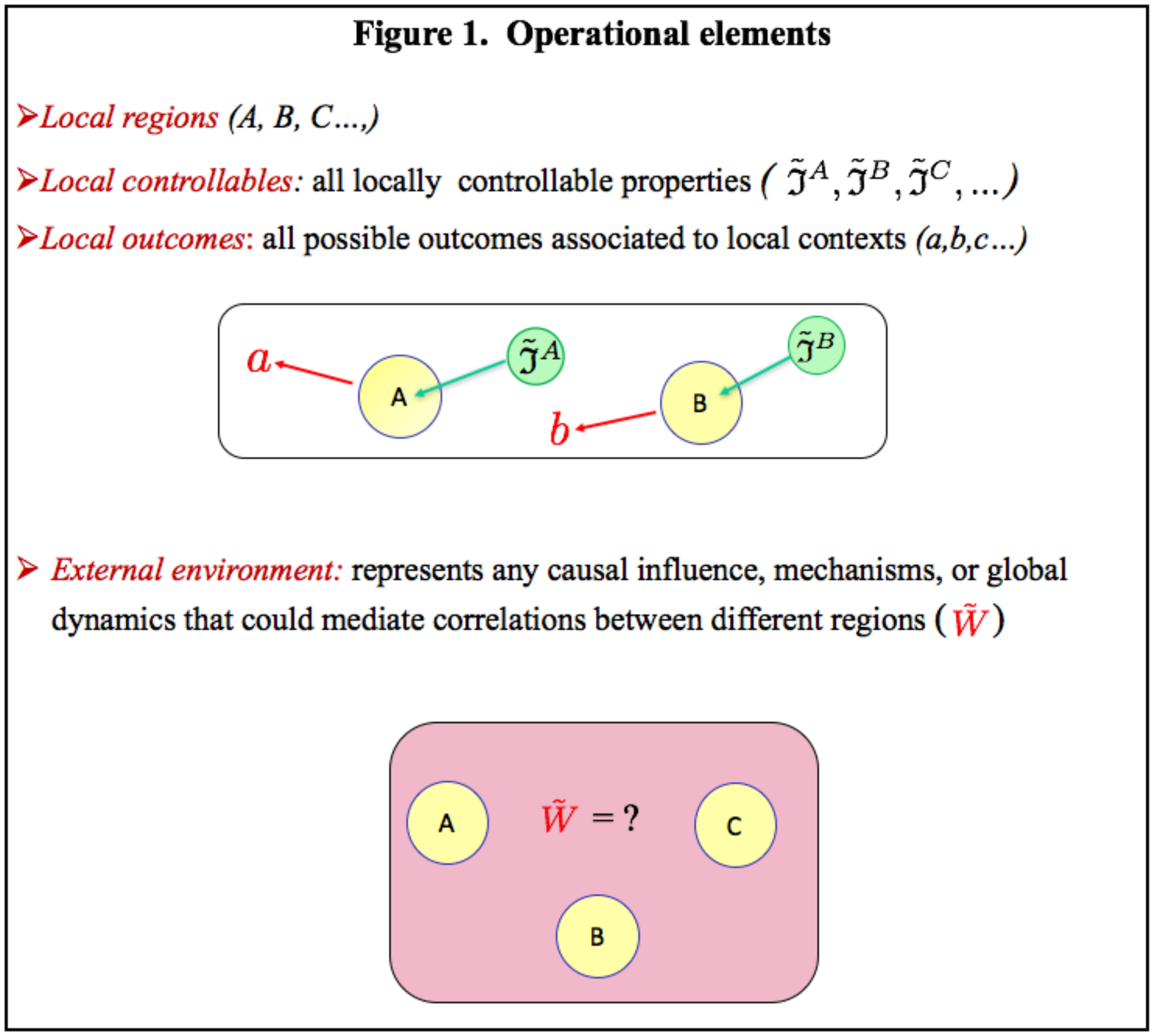}%
\caption{\textbf{Operational primitives.}}%
\label{region}%
\end{figure}

We can thus describe an experiment by a set of regions, outcomes, local controllables and an environment. If we consider a particular run of an experiment there will in general be a collection of outcomes that occur, one for each local region. One can associate a joint probability to this set of outcomes and empirically verify probability assignments for each possible set of outcomes. An operational model for such an experiment allows one to calculate expected probabilities: 
\begin{equation}
p(a, b, c,\dots| \tilde{\mathcal{\mathfrak{I}}}^A, \tilde{\mathcal{\mathfrak{I}}}^B, \tilde{\mathcal{\mathfrak{I}}}^C,\dots, \tilde{W}).
\end{equation}

The operational model thus specifies a distribution over outcomes for local controllables  $\tilde{\mathcal{\mathfrak{I}}}^A, \tilde{\mathcal{\mathfrak{I}}}^B, \dots$, and a shared environment $\tilde{W}$, Fig.~\ref{region}. Note that it should be possible to have ignorance over part of the environment and characterise this accordingly using the operational model. More explicitly, if $\tilde{\xi}$ represents the part of the environment about which we are ignorant, then the operational probabilities given the known part of the environment are obtained by marginalising over $\tilde{\xi}$:

\begin{align}\label{marginal}
p(a, b, c,\dots| \tilde{\mathcal{\mathfrak{I}}}^A, \tilde{\mathcal{\mathfrak{I}}}^B, \tilde{\mathcal{\mathfrak{I}}}^C,\dots, \tilde{W}) =& \int d\tilde{\xi} p(a, b, c,\dots, \tilde{\xi}| \tilde{\mathcal{\mathfrak{I}}}^A, \tilde{\mathcal{\mathfrak{I}}}^B, \tilde{\mathcal{\mathfrak{I}}}^C,\dots, \tilde{W})\\ \nonumber
=& \int d\tilde{\xi} p(a, b, c,\dots| \tilde{\mathcal{\mathfrak{I}}}^A, \tilde{\mathcal{\mathfrak{I}}}^B, \tilde{\mathcal{\mathfrak{I}}}^C,\dots, \tilde{W},\tilde{\xi}) p( \tilde{\xi}| \tilde{W}),
\end{align}
where the second equality comes from the assumption that the local controllables are uncorrelated with the environment. 

As a concrete example, $\tilde{W}$ can describe the axis along which a spin-$\frac{1}{2}$ particle is prepared, while $\tilde{\xi}$ represents whether the spin is prepared aligned or anti-aligned with that axis.\footnote{Here (and again in Section~\ref{quantModel}), we take for simplicity a scenario with a single region where a measurement is performed, so the specification of a process is equivalent to the specification of a state. More generally, the variables $W$ and $\xi$ could describe quantum channels, quantum networks, or more general quantum processes.} The marginal \eqref{marginal} then describes a scenario where there is some probabilistic uncertainty of the spin's direction i.e. which value of $\xi$ occurs in any given run. Note that, for the particular case $p( \tilde{\xi}| \tilde{W})=\frac{1}{2}$, we obtain the maximally mixed state irrespective of the axis, making the variable $\tilde{W}$ redundant. Such redundancies can be taken into account via operational equivalences.

\section{Operational equivalences}\label{opEquiv}
We next characterise the appropriate operational equivalences in order to define our ontological model. Notationally, we omit the `tilde' for each equivalence class.
\subsection{Events}
We say that a pair composed of an outcome and the respective local controllable $(a, \tilde{{\mathfrak{I}}}^A)$ is operationally equivalent to the pair $(a', \tilde{{\mathfrak{I}}}'^A)$ if the joint probabilities for $a, b, c,\dots$ and $a', b, c, \dots$ are the same for all possible outcomes and local controllables in the other regions $B, C,\dots,$ and for all environments $\tilde{W}$.
\begin{equation}
 p(a,b,c,\dots | \tilde{\mathcal{\mathfrak{I}}}^A, \tilde{\mathcal{\mathfrak{I}}}^B,\dots\tilde{W}) = p(a',b,c,\dots | \tilde{\mathcal{\mathfrak{I}}}'^A, \tilde{\mathcal{\mathfrak{I}}}^B,\dots\tilde{W}),
\end{equation}
\begin{equation*}
\forall (b,c,\dots,\tilde{\mathcal{\mathfrak{I}}}^B, \tilde{\mathcal{\mathfrak{I}}}^C,\dots, \tilde{W}).
\end{equation*}

We denote an equivalence class of such pairs of outcomes and local controllables as an \emph{event}:
\begin{equation}\label{events}
M^A = [(a, \tilde{\mathcal{\mathfrak{I}}}^A)].
\end{equation}

\subsection{Instruments}

We define an instrument as the list of \emph{possible events} for a local controllable $ \tilde{\mathcal{\mathfrak{I}}}^A $, where an event $M^A = [(a, \tilde{\mathcal{\mathfrak{I}}}^A)]$ is possible for $\tilde{\mathcal{\mathfrak{I}}^A}$ if

\begin{equation}
 p(a,b,c,\dots | \tilde{\mathcal{\mathfrak{I}}}^A, \tilde{\mathcal{\mathfrak{I}}}^B,\dots\tilde{W}) \neq 0,
\end{equation}
 for some
\begin{equation*}
 (b,c,\dots,\tilde{\mathcal{\mathfrak{I}}}^B, \tilde{\mathcal{\mathfrak{I}}}^C,\dots, \tilde{W}).
\end{equation*}

We say that $\tilde{\mathcal{\mathfrak{I}}}^A$ is equivalent to $\tilde{\mathcal{\mathfrak{I}}}'^A$ if they define the same list of possible events and we denote the equivalence class ${\mathcal{\mathfrak{I}}}^A := [\tilde{\mathcal{\mathfrak{I}}}^A] \equiv \{M_1^A, \dots, M_n^A\}$.
Note that our definition allows distinct instruments to share one or more events. Note also, our definition implies that the probability for an event doesn't depend on the particular instrument $\mathcal{\mathfrak{I}}$, once we assume the event is possible given the instrument. This property we call \emph{operational instrument equivalence}.\footnote{In other work, where we are not concerned with the possibility of non-contextual hidden variable theories, we refer to this property as instrument non-contextuality~\cite{shrapnel2017}. Here we reserve the term non-contextuality to refer to an ontological model.}

\subsection{Process}
The process captures those physical features responsible for generating the joint statistics for a set of events, independently of the choice of local instruments. A process is defined as an equivalence class of environments, $W:= [\tilde{W}]$, where
$\tilde{W}$ is equivalent to $\tilde{W}' $, if
\begin{equation}
p(a,b,c,\dots | \tilde{\mathcal{\mathfrak{I}}}^A, \tilde{\mathcal{\mathfrak{I}}}^B,\dots,\tilde{W}) = p(a,b,c,\dots | \tilde{\mathcal{\mathfrak{I}}}^A, \tilde{\mathcal{\mathfrak{I}}}^B,\dots\tilde{W'}),
\end{equation}
\begin{equation*}
\forall (a, b,c,\dots,\tilde{\mathcal{\mathfrak{I}}}^B, \tilde{\mathcal{\mathfrak{I}}}^C,\dots, ).
\end{equation*}

A simple example is the spatio-temporal ordering of regions. It is clear that the operational statistics of events in regions A and B can be different for the following two causal orderings: (i) A is before B, (ii) B is before A; thus the respective environments, $\tilde{W}_{(i)}$ and $\tilde{W}_{(ii)}$, will not be equivalent. On the other hand, for certain experiments we would not expect any difference in statistics for a simple rotation of the whole experiment by 45 degrees; these two environments will be represented by the same process $W$.

The above equivalences allow us to define a joint probability distribution over the space of \emph{events} (rather than outcomes) conditioned on \emph{instruments} (rather than local controllables) and the \emph{process} (rather than the environment). As discussed above, this distribution satisfies operational instrument equivalence, which means that the joint probability for a set of events is either zero or independent of the respective instruments. Therefore, it can be expressed in terms of a \emph{frame function} $f_{W}$ that maps events to probabilities and is normalised for each instrument:
\begin{equation}
p(M^A, M^B,\dots|\mathfrak{I}^A, \mathfrak{I}^B,\dots, W) =  f_W(M^A, M^B,\dots) \prod_{X=A,B,\dots}  \chi_{\mathfrak{I}^X}(M^X),
\end{equation}
where, for a set $S$, $\chi_S$ is the indicator function, $\chi_S(s)=1$ for $s\in S$ and $\chi_S(s)=0$ for $s\not\in S$. Note that the indicator functions are necessary to make the whole expression a valid probability distribution, normalised over the \emph{entire} space of events. 
Furthermore, and in contrast to similar expressions involving POVMs, the dependency on the instruments is crucial to allow for causal influence across the regions: Integrating over the events of, say, region $A$, can result in a marginal distribution that still depends on $A$'s instrument and displays signalling from $A$ to other regions. However, the fact that the dependency on the instruments is solely through the indicator functions tells us that the causal relations can be attributed to the particular events realised in each experimental run, rather than to the whole instruments (which include the specification of events that did not happen). In other words, the event ``screens off'' the instrument: once the event in a local region is known, further knowledge of the instrument does not allow for any better prediction about events in other regions.

 \section{Ontological model}\label{ontModel}

The purpose of an ontological model is to introduce possible elements of reality. Typically, one assumes that the ontology is encoded in a ``state'', representing the physical properties of a system at a given time.  Here we shift the focus from states to more general properties of the environment that are responsible for mediating correlations between regions. We represent the collection of all such properties by a single variable $\omega$, named the \emph{ontic process}. We wish to clarify at this point that our ontic process captures the physical properties of the world that remain invariant under our local operations. That is, although we allow local properties to \emph{change} under specific operations, we wish our \emph{ontic process} to capture those aspects of reality that are independent of this probing. The interpretation of ontic processes and the relation with the usual notion of ontic states can be seen via the examples of the following section.

Our ontological model specifies a joint probability for a set of outcomes, one at each local region, given the ontic process, the environment, and the set of local controllables. This joint probability reduces to the operational joint probability when the value of the ontic process is unknown:
\begin{equation}
p(a,b,c,\dots, | \tilde{\mathcal{\mathfrak{I}}}^A, \tilde{\mathcal{\mathfrak{I}}}^B,\dots\tilde{W}) = \int d\omega p(a,b,c,\dots, \omega | \tilde{\mathcal{\mathfrak{I}}}^A, \tilde{\mathcal{\mathfrak{I}}}^B,\dots\tilde{W}).
\end{equation}

There are three natural assumptions one might require of an ontological model defined according to these operational equivalences:

\begin{assumption}
\textbf{$\omega$-mediation}
The ontic process mediates all the correlations between regions, thus $\omega$ screens off outcomes from the environment, and we have: 
\begin{equation}
p(a,b,c,\dots| \tilde{\mathcal{\mathfrak{I}}}^A, \tilde{\mathcal{\mathfrak{I}}}^B,\dots,\tilde{W}) = \int d\omega p(a,b,c,\dots | \omega, \tilde{\mathcal{\mathfrak{I}}}^A, \tilde{\mathcal{\mathfrak{I}}}^B,\dots) p(\omega|\tilde{W}).
\end{equation}

\end{assumption}

\begin{assumption}
\textbf{Instrument non-contextuality.}
Operationally indistinguishable pairs of outcomes and local controllables should remain indistinguishable at the ontological level.
That is, for operationally equivalent pairs $(a, \tilde{\mathcal{\mathfrak{I}}}^A), (a', \tilde{\mathcal{\mathfrak{I}}}'^A)$,
\begin{equation}
p(a,b,c,\dots, | \omega, \tilde{\mathcal{\mathfrak{I}}}^A, \tilde{\mathcal{\mathfrak{I}}}^B,\dots) =  p(a',b,c,\dots, | \omega, \tilde{\mathcal{\mathfrak{I}}}'^A, \tilde{\mathcal{\mathfrak{I}}}^B,\dots),
\end{equation}
\begin{equation*}
\forall (b,c,\dots,\tilde{\mathcal{\mathfrak{I}}}^B,\dots), \forall \omega.
\end{equation*}
which means that we can define a probability distribution on the space of events, conditioned on instruments and on the ontic process, in terms of a \emph{frame function} $f_\omega$, such that:
\begin{equation}
p(M^A, M^B,\dots| \omega,\mathfrak{I}^A, \mathfrak{I}^B,\dots,) = \prod_{X}  \chi_{\mathfrak{I}^X}(M^X) f_\omega(M^A, M^B\dots),
\end{equation}
where $\chi$ is the indicator function, $\chi_X(x)=1$ for $x\in X$ and $\chi_X(x)=0$ for $x\not\in X$, and  $f_\omega$ maps events to probabilities:
\begin{equation}
 f_\omega(M^A, M^B, \dots) \in [0,1],
\end{equation}
and is normalised for each set of events that corresponds to a particular instrument:
\begin{equation}
{\sum_{\substack{M^A \in \mathfrak{I}^A\\{M^B \in \mathfrak{I}^B}\\{M^C \in \mathfrak{I}^C}\\\dots}}}f_\omega(M^A, M^B,M^C, \dots) = 1.
\end{equation}
\end{assumption}

\begin{assumption}
\textbf{Process non-contextuality.}
\\
For operationally equivalent processes $\tilde{W}, \tilde{W}'$ the assumption of process non-contextuality implies:
\begin{equation}
p(\omega|\tilde{W}) = p(\omega|\tilde{W}'),
\end{equation}
and we can define a function $g_W(\omega)$ that maps ontic processes to probabilities, given each process $W$:
\begin{equation}
g_W(\omega) = p(\omega|\tilde{W}),\quad W=\left[\tilde{W}\right],
\end{equation}
that is normalised for all $\omega$:
\begin{equation}
\int d \omega~g_W (\omega) = 1.
\end{equation}
\end{assumption}
For an ontological model that satisfies the above three assumptions, the operational probability can now be expressed in terms of events, instruments and processes as:
\begin{equation}
p(M^A, M^B,\dots|\mathfrak{I}^A, \mathfrak{I}^B,\dots, W) 
 = \prod_{X=A,B,\dots} \chi_{\mathfrak{I}^X}(M^X) \int d\omega~f_\omega(M^A, M^B, \dots) g_W(\omega).
\end{equation}

Although ontic states, as they are usually understood, are not represented explicitly in our framework, they are not excluded. In the following section we present three examples to illustrate how such ontic states, with or without retrocausality, can be represented in our model.

\section{Examples}\label{examples}

\subsection{Deterministic, classical models}
\subsubsection{Causally-ordered models}
\begin{figure}[ht]%
\subfloat[\label{figure2a}]{
\includegraphics[width=0.5\columnwidth]{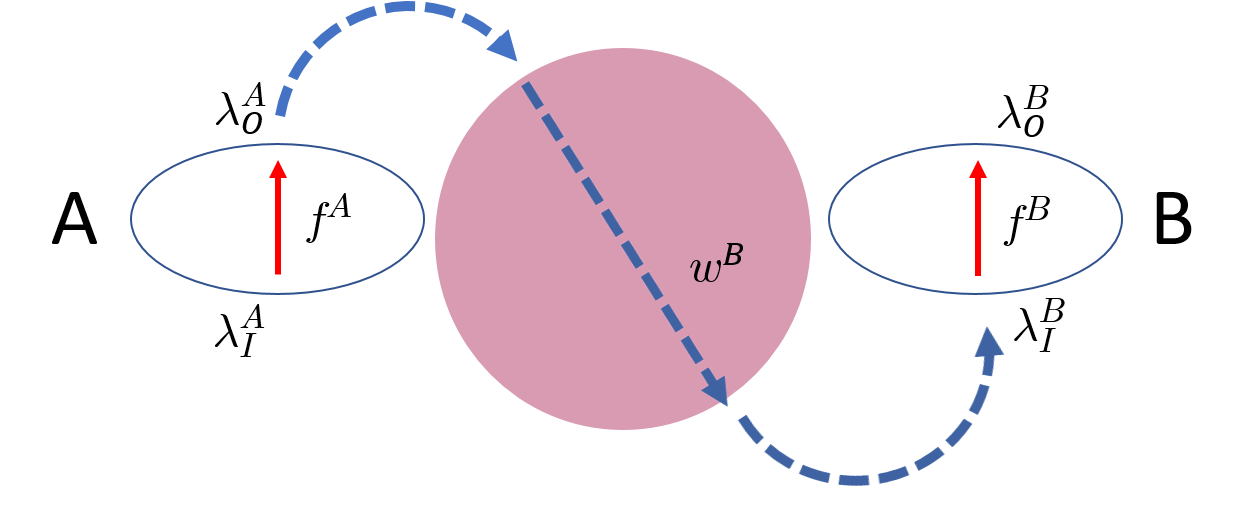}}
\quad
\subfloat[\label{figure2c}]{
\includegraphics[width=0.5\columnwidth]{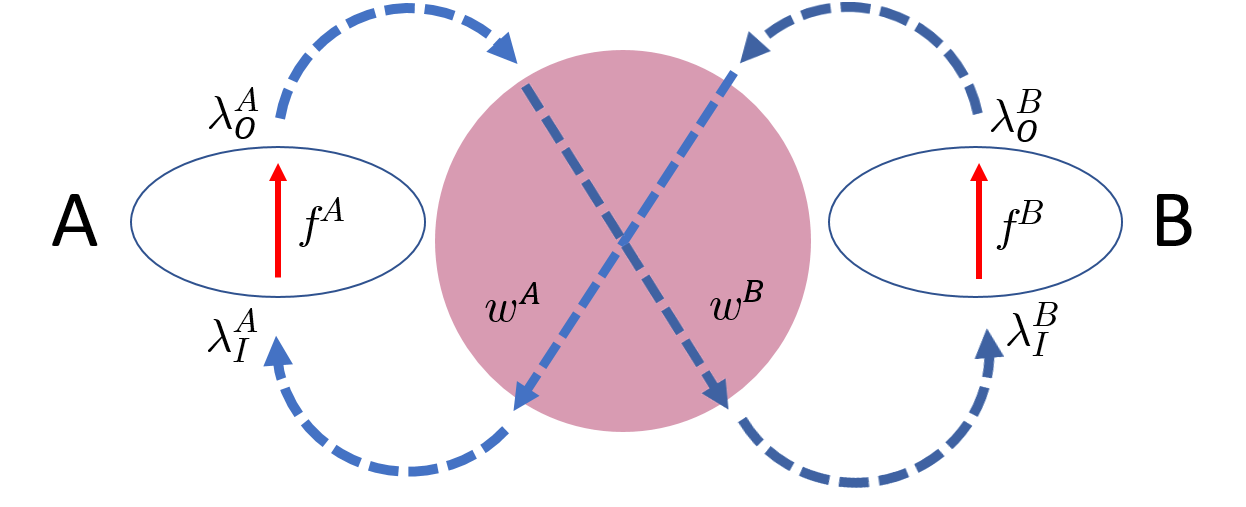}
}
\caption{\textbf{ Classical process with ontological interpretation.}{ (a) We assign input states $\lambda_I^A$ and $\lambda_I^B$ to the past boundaries of $A$ and $B$, respectively, and output states $\lambda_O^A$ and $\lambda_O^B$ to the respective future boundaries. A deterministic local operation is a function $f^X$ that maps the input state of each region to the corresponding output. (b) An example of ontic process is one describing classical closed time-like curves, defined by a pair of functions $\omega=\left(w^A, w^B\right)$, where $\lambda_O^B \mapsto w^A\left(\lambda_O^B\right) = \lambda_I^A$ and similarly for $w^B$. }}
\end{figure}

As a first example, let us consider a classical, deterministic scenario (without retrocausality) with two regions, $A$ in the past of $B$, each delimited by a past and future space-like boundary, see Fig.~\ref{figure2a}. For a classical system, we can assign input states $\lambda_I^A$ and $\lambda_I^B$ to the past boundaries of $A$ and $B$, respectively, and output states $\lambda_O^A$ and $\lambda_O^B$ to the respective future boundaries. As measurements can be performed without disturbance on a classical system, we associate the input state in each region with the respective measurement outcome: $a\equiv \lambda_I^A$ and $b \equiv \lambda_I^B$.
As local controllables we take deterministic local operations, defined as functions $f^X$ that map the input state of each region to the corresponding output:
\begin{equation}
 \lambda_I^X\mapsto\lambda_O^X=f^X\left(\lambda_I^X\right),
 \end{equation} where X denotes the respective local region, $A$ or $B$. 
Assuming ordinary dynamical laws, the input state at $B$ can depend on the output at $A$ through some function:
 \begin{equation}\lambda_O^A \mapsto \lambda_I^B = w^B\left(\lambda_O^A\right).
 \end{equation} The input state at $A$, on the other hand, does not depend on $B$, and thus has to be specified as an independent environment variable. The ontic process for this model is thus identified with the pair
 \begin{equation}
 \omega = \left(\lambda^A_I, w^B\right).
 \end{equation}
  Indeed, knowing $\omega$ and the choice of local operations is sufficient to fully determine the measured outcomes: 
  \begin{equation}
  a=\lambda^A_I, b= w^B\left(f^A\left(\lambda^A_I\right)\right).
  \end{equation} 

As the model is fully deterministic, and we have not introduced any redundant variables, there are no non-trivial equivalence classes. Explicitly, an event in region $A$ (and similarly for $B$) is given by the pair $\left(a,f^A\right)$, or equivalently by the input-output pair 
\begin{equation}
\lambda^A := \left(\lambda^A_I,\lambda^A_O = f^A\left(\lambda^A_I\right)\right), 
\end{equation}
 while the instrument is given by the collection of events given a choice of operation, 
 \begin{equation}
 \mathfrak{I}^A = \left\{\left(\lambda_I^A,\,f^A\left(\lambda_I^A\right)\right)\right\}_{\lambda_I^A},
 \end{equation}
 which is just to say the instrument can be identified with the function $f^A$. We see in this example that the ontology, as traditionally understood, lies in the event variables $\lambda$. These variables are \emph{not} independent of the local controllables, because the event at $B$ can depend on the operation performed at $A$. However, there is still an aspect of the ontology that does not depend on the operations: the initial state $\lambda^A_I$ and the functional relation $w^B$. It is this invariant aspect of the ontology that we call a process.

\subsubsection{Time-travelling classical systems}
General Relativity allows for space-time geometries with closed time-like curves, where a system can travel back in time and interact with its past self \cite{morris1988wormholes}, thus providing physically-motivated examples of scenarios that defy ordinary forward causality. Notably, qualitative analogies between quantum phenomena and classical time-travelling systems have been suggested~\cite{Durand2002}, making the latter an interesting test-bed for generalised ontological models.

The example in the previous subsection can be readily generalised to a deterministic model of classical system near closed time-like curves by allowing the input state at $A$ to depend on $B$ through some function $\lambda_O^B \mapsto \lambda_I^A = w^A\left(\lambda_O^B\right)$. The process is now given by two functions, $\omega\equiv\left(w^A, w^B\right)$, Fig.~\ref{figure2c}, with the causally-ordered case recovered when one of the two is a constant. Compatibility with arbitrary local operations imposes constraints on the function $w^A, w^B$ and, in the two-region case, it turns out that one of them has in fact to be constant~\cite{Baumeler2016, baumeler2017reversible}. However, for three or more regions, it is possible to find deterministic processes, with no constant component, that are still consistent with arbitrary local operations\footnote{The incompatibility of such processes with an underlying causal order can be demonstrated rigorously by showing they can be used to violate causal inequalities~\cite{baumeler14}, device-independent constraints on probabilities imposed by a definite causal order~\cite{oreshkov12, Branciard2016}.}.

Also in this case, the observed outcomes are fully determined once the process and the local operations are specified, as the unique fixed points $a\equiv \lambda_I^A, b\equiv \lambda_I^B,\dots$, of the function obtained by composing the process  $\omega=\left(w^A, w^B,\dots\right)$ with the operations $\left(f^A,f^B,\dots\right)$ (see Ref.~\cite{baumeler2017reversible} for more details). Crucially, in this case the events in each region can depend on the choice of operation in all regions, $\lambda^A=\lambda^A(f^A,f^B,\dots)$. Thus, from the perspective of ordinary ontological models, time-travelling systems appear contextual, since it is impossible to assign a ``state'' to any region independently of the operations. Nonetheless, the relation between events, captured by the process, does not depend on the operations. Thus, following the terminology introduced here, models such as the above are both instrument and process non-contextual. (As in the previous causally-ordered example, there are no non-trivial equivalence classes, so non-contextuality is straightforward.)

More general models of classical closed time-like curves might impose restrictions on the accessible local operations\footnote{ A constraint on the accessible operations is often invoked to solve ``paradoxes'', such as a time-traveller killing their past self. Although classical studies of closed time-like curves do not support the need for this type of restriction~\cite{Friedman:1990ja, Echeverria:1991ko, Lossev1992, Novikov1992, Mikheeva1993}, it might be necessary in a general theory. Such a restriction on an agent's actions is sometimes interpreted as a violation of ``free will''. This worry is however misplaced, since an agent can still be (or fail to be) free to perform all the physically available operations. A different set of operations would simply represent a deviation from classical physics in the local region where the agent acts.}. Even more generally, one can consider models where instruments are not associated with local input-to-output functions but with more general sets of input-output pairs, $\mathfrak{I}^A=\Lambda^A\subsetneq \Lambda^A_I\times\Lambda^A_O$, where $\Lambda^A_{I (O)}$ is the state space associated with the past (future) boundary of the local region. In such models, a choice of instrument selects which pairs of input-output states are \emph{possible}, while a deterministic process would determine, given all choices of instruments, which pairs are actually realised. Thus, in such models both the state in the past and in the future of a local region depend on the choice of instrument, thus again they are necessarily contextual from the point of view of traditional ontological models. Yet, they remain instrument and process non-contextual as long as deterministic processes are considered.

In the above deterministic examples $\omega$-mediation is satisfied trivially, because ontic and operational processes coincide. This can be generalised to situations where we have only partial knowledge about the environment. For example, we might not have full knowledge of the initial state, but only know the temperature $T$ of a thermal bath from which the state is extracted; or the system might get coupled to some external environment during the evolution from one region to another. In all cases, we end up with partial knowledge of the ontic process, expressed by some probability $p(\omega|W)$ where $W$ represents all relevant accessible information about the environment (the temperature of the bath or other noise parameters). The resulting probabilistic operational model naturally satisfies the property of $\omega$-mediation, because knowing the temperature or noise parameters does not provide more information than already encoded in the ontic process, namely in the underlying microstates and functional relations. 

Note also that our construction of an ontological model respects the mobility of the boundary between local instruments and processes that one sees in ordinary applications of quantum theory. As a simple example, consider a preparation $P$ of a quantum system, followed by a measurement $M$. This can be modelled in three different ways: (i) with P as part of the environment $\tilde{W}$, and $M$ as an instrument associated to a single local region, (ii) with $P$ and $M$ as instruments in two distinct local regions, and $\tilde{W}$  capturing both a channel between preparation and measurement, plus any additional information about the environment, or  (iii) with both $P$ and $M$ characterising the instruments in a single local region and all other information about the environment modelled as $\tilde{W}$. For classical processes characterised as causal models, such a shift in perspective is formalised by the notion of ``latent variables''\cite{Pearlbook}. An analogue notion of "latent laboratories" exists for quantum processes characterised as quantum causal models, and this formal structure likewise characterises the mobility of the boundary to which we refer\cite{costa2016}.

\subsubsection{All-at-once stochastic models}
In the above examples of time-travelling systems, the ontic process (or at least certain aspects of it) can be understood as describing the dynamical evolution of systems between regions. Some retrocausal approaches attempt to provide an ontology for quantum mechanics that does not rely on any dynamical process; rather, one should consider all relevant events in space-time ``at once''. The appearance of quantum probabilities is then justified by the fact that the information available at a given time is not sufficient to fully determine the state of the system at all times (with the missing information possibly contained in some unknown boundary condition in the future). Our framework naturally captures all such models, because an ontic process need not be interpreted as a transformation: it simply represents the rule generating all relevant events given the local operations. 

An instructive example is a toy model by Wharton~\cite{Wharton2014}, which represents a space-time scenario as a system in thermal equilibrium, with events at different space-time locations represented as states at different points in space. While having a clear ontological interpretation, this model offers qualitative analogies with quantum interference and, when analysed from an ordinary time-evolution perspective, displays an apparent contextuality. We show in detail in appendix \ref{toywharton} how (a generalisation of) Wharton's model fits within our framework and satisfies the requirements of $\omega$-mediation and instrument and process non-contextuality.

The above three examples illustrate that it is indeed easy to represent many possible physical scenarios via ontological models that are both instrument and process non-contextual. Given the exotic nature of the latter two examples, it seems plausible that one could also produce such a model to explain quantum correlations. In the following sections we prove that this is not the case.

\section{Quantum models}\label{quantModel}

If one assumes that the results of experiments in local regions accord with quantum mechanics, then events can be associated with \emph{completely positive trace-non-increasing} (CP) maps $\mathcal{M}^A :A_I\rightarrow A_O$, where input and output spaces are the spaces of linear operators over input and output Hilbert spaces of the local region, $A_I\equiv\lin({\cal H}^{A_I})$, $A_O\equiv \lin({\cal H}^{A_O})$ respectively~\cite{chuang00}. Each set ${\mathfrak I}^A$ of  CP maps that sums to a completely positive trace preserving (CPTP) map is a \emph{quantum instrument}~\cite{davies70}:
\begin{equation}
\tr \left[ \sum_{\mathcal{M}^A\in {\mathfrak I}^A} \mathcal{M}^A(\rho)\right] =  \tr(\rho).
\label{instrument}
\end{equation}
An instrument thus represents the collection of all possible events that can be observed given a specific choice of local controllable.

Given these definitions of events and instruments, one can predict the joint probability over possible events using a generalised form of the Born rule:
\begin{align}
p(M^A, M^B,\dots|\mathfrak{I}^A, \mathfrak{I}^B,\dots, W)  =&f_W(M^A, M^B,\dots)
\prod_{X}  \chi_{\mathfrak{I}^X}(M^X),\\ \label{born}
f_W(M^A, M^B\dots) =& \tr \left[(M^A \otimes M^B \otimes\dots) W\right],
\end{align}
where $M^A, M^B\dots$ are the Choi-Jamio{\l}kowski representations of the local CP maps associated to particular events, and $W$ is a positive, semi-definite operator associated to the relevant process~\cite{gutoski06, chiribella09b, oreshkov12}. We call $W$ the \emph{process matrix}, using the terminology of Ref.~\cite{oreshkov12}.

It is possible to derive this trace rule for probabilities by assuming linearity~\cite{oreshkov12}, or alternatively one can \emph{derive} linearity (and the trace rule) from the assumption of operational instrument equivalence alone~\cite{shrapnel2017}. The significance of this latter derivation is that the condition of operational instrument equivalence is formally identical to that of instrument non-contextuality, with the only difference that the latter includes the ontic process. Therefore,
for each ontic process $\omega$, the corresponding frame function can be expressed as:

\begin{equation}\label{framenoncontextual}
f_\omega(M^A, M^B,\dots) = \tr\left[ \sigma(\omega) \mathbf{M}\right],
\end{equation}
where we introduced the short-hand notation $\mathbf{M} \equiv M^A \otimes M^B\otimes\dots$ and $\sigma(\omega)$ is a process matrix~\cite{shrapnel2017}. We now wish to show that the function $g_W(\omega)$ that features in our ontological model, under the assumption of process non-contextuality, can be represented as
\begin{equation}
\label{dualframe}
g_W(\omega) = \tr \left[\eta(\omega)W\right],
\end{equation}
where $\{\eta(\omega)\}_{\omega \in \Omega}$, $\Omega$ being the set of ontic processes, is a quantum instrument. 

It is common in non-contextuality no-go theorems (as well as in the process matrix formalism) to assume preservation of probabilistic mixtures as an assumption that is independent of the assumption of non-contextuality. Here we rather derive it from our assumption of process non-contextuality. Consider two classical variables  $\xi$, $W$ used to describe the process, where we already take operational equivalences into account. Following the earlier example, we can think of $W$ as describing a cartesian axis, while $\xi$---the aspect of the process about which we are ignorant---describes whether a spin-$\frac{1}{2}$ particle is prepared aligned or anti-aligned to this axis. The operational probabilities given $W$, and the corresponding decomposition for ontological probabilities, are obtained by marginalisation:
\begin{align}\label{marginalprocess}
p&(M^A,M^B,\dots|\mathfrak{I}^A, \mathfrak{I}^B,\dots, W)\\ \nonumber
=&~\int\! d\xi\,d\omega\, p(M^A,M^B,\dots|\omega, \mathfrak{I}^A, \mathfrak{I}^B,\dots, W,\xi)p(\omega| \mathfrak{I}^A, \mathfrak{I}^B,\dots,W, \xi)p(\xi|\mathfrak{I}^A, \mathfrak{I}^B,\dots,W)\\ \nonumber
=&~\int\! d\xi\,d\omega\, p(M^A,M^B,\dots|\omega, \mathfrak{I}^A, \mathfrak{I}^B,\dots)p(\omega | W, \xi) p(\xi|W),
\end{align}
where, in the last identity, we use the fact that $p(\omega | W, \xi)$ does not depend on the local controllables (and thus on the instruments) due to the assumption of $\omega$-mediation; and $p(\xi|W)$ is due to our assumption that the environment and local controllables (and thus process and instruments) are uncorrelated. Additionally, due to $\omega$-mediation, we no longer need to condition the $M^A,M^B,\dots$ directly on $W$ and $\xi$.

Now let us write $W_{\xi}$ for the process corresponding to the pair $W, \xi$. We have
\begin{equation}
\label{convexity}
g_{\int d{\xi}~W_{\xi}p({\xi}|W)}(\omega)= g_W(\omega) =  p(\omega|W) 
= \int d{\xi}~ g_{W_{\xi}} (\omega) p({\xi}|W), 
\end{equation}
thus $g_W(\omega)$ is convex-linear in $W$. The first identity in Eq.~\eqref{convexity} comes from the fact that probabilistic mixtures of quantum processes are represented as convex combinations, thus $W = \int d{\xi}~W_{\xi} p({\xi}|W)$. This in turn is a consequence of the trace formula for operational quantum probabilities (which is itself a consequence of operational instrument equivalence):
\begin{align}
p(M^A, M^B,\dots|\mathfrak{I}^A, \mathfrak{I}^B,\dots, W) 
&= \tr \left[(M^A \otimes M^B \otimes\dots) W \right] \prod_{X}  \chi_{\mathfrak{I}^X}(M^X)\\ \nonumber
&= \int d{\xi }~p(M^A, M^B,\dots|\mathfrak{I}^A, \mathfrak{I}^B,\dots, W,\xi) p({\xi}|W) \prod_{X}  \chi_{\mathfrak{I}^X}(M^X) \\
&=  \tr\left[ (M^A \otimes M^B \otimes\dots)\int d{\xi }~W_{\xi } p({\xi}|W) \right] \prod_{X}  \chi_{\mathfrak{I}^X}(M^X)
\end{align}
for all CP maps $M^A, M^B,\dots$

Using standard linear-algebra arguments, $g_W(\omega)$ can be extended to a linear function over $W$, leading to the representation \eqref{dualframe}, $g_W(\omega)= \tr\left[ \eta(\omega)W\right]$. Positivity and normalisation of probabilities then imply

\begin{align}
g_W(\omega) \geq 0 &\Rightarrow \eta(\omega) \geq 0 \quad \forall \omega, \\ \label{etanormalisation}
\int d\omega  g_W(\omega) = 1 &\Rightarrow \tr\left[ \int d\omega \eta(\omega) W \right]= 1 \quad \forall\, W. 
\end{align}
Operators $\eta(\omega)$ as defined above can be understood as the Choi representation of CP maps that sum up to a trace preserving map, namely $\left\{\eta(\omega)\right\}_{\omega\in \Omega}$ defines an instrument. In general, the CP maps $\eta(\omega)$ do not have to factorise over the separate regions, therefore it might not be possible to interpret them as local operations. This is not an obstacle, as such an interpretation is not required for the rest of the argument.

\section{A quantum contradiction}\label{contradiction}

To summarise the results so far, we have an operational rule for the predictions of the joint probabilities of outcomes according to quantum theory:
\begin{equation}
p(M^A, M^B,\dots|\mathfrak{I}^A, \mathfrak{I}^B,\dots, W)  =\prod_{X} \chi_{\mathfrak{I}^X}(M^X)\tr\left[\mathbf{M}\,W\right] .
\end{equation} 
We also have an ontological model for predicting the joint probabilities under the assumptions of $\omega$-mediation, instrument non-contextuality and process non-contextuality:
\begin{equation}\label{inframes}
p(M^A, M^B,\dots|\mathfrak{I}^A, \mathfrak{I}^B,\dots, W)  = \prod_{X}  \chi_{\mathfrak{I}^X}(M^X) \int d\omega f_\omega(M^A, M^B, \dots) g_W(\omega),
\end{equation}
which given the results of the last section, becomes:
\begin{equation}
p(M^A, M^B,\dots|\mathfrak{I}^A, \mathfrak{I}^B,\dots, W) =  \prod_{X} \chi_{\mathfrak{I}^X}(M^X)\int d\omega \left[ \tr \sigma(\omega) \mathbf{M}\right] \left[\tr \eta(\omega)W\right].
\end{equation}
If this accords with quantum predictions then we should have:
\begin{equation}\label{quantont}
\tr\left[\mathbf{M}\,W\right] =  \int d\omega~\left[ \tr \sigma(\omega) \mathbf{M}\right] \left[\tr \eta(\omega)W\right]~\forall M, W.
\end{equation}

It has been noted \cite{spekkens08} that a decomposition of the form \eqref{inframes} is akin to the expression of expectation values in terms of quasi-probability distributions~\cite{Wigner1932, Scully1997}. However, the non-contextuality assumptions force both $f_\omega$ and $g_W$ to be ordinary, positive probability distributions. It is well known that quantum expectation values cannot be expressed in such a way. It is however instructive to consider an explicit contradiction within the present process framework.

From \eqref{quantont},
\begin{eqnarray}
\tr\left[\mathbf{M}\, W\right] =  \tr\left[\mathbf{M} \int d\omega~\sigma(\omega)~g_W(\omega)\right] \quad \forall \mathbf{M} \\
\rightarrow W= \int d\omega~\sigma(\omega)~g_W(\omega),
\label{ned}
\end{eqnarray}
which follows from the fact that $\mathbf{M}$ span a complete set of the joint linear space $A^I\otimes A^O\otimes B^I\otimes B^O,\dots$

Eq.~\eqref{ned} tells us that $W$ is a convex mixture of the operators $\sigma(\omega)$. If $W$ is extremal, namely if it cannot be decomposed into a non-trivial convex combination of other processes, then $W \propto \sigma(\omega)$ for $g_W(\omega) \neq 0$. Denoting the support of $g_W$ by $\Omega_W$, i.e., $\omega\in \Omega_W \Leftrightarrow g_W(\omega) \neq 0$, we have $W \propto \sigma(\omega)$ $\forall \omega \in \Omega_W$ for an extremal $W$.

Consider now a process $W$ that can be decomposed into two distinct mixtures of two sets of extremal processes $W_j$ and $W'_k$ (we take discrete sets for simplicity):
\begin{equation} 
W= \sum_j q_j W_j = \sum_k p_kW'_k.
\end{equation}
Since $g_W$ is convex-linear in $W$, we have $g_W= \sum_j q_j g_{W_j}$.
This means that, for every $\omega \in \Omega_W$, there must be a $j$ such that $g_{W_j}(\omega) \neq 0$. In other words, $\Omega_W=\bigcup_j \Omega_{W_j}$. By a similar argument, we have that $\Omega_W=\bigcup_k \Omega_{W'_k}$. We thus see that each convex decomposition of $W$ into distinct extremal processes corresponds to a partition of $W$'s support into the extremal processes' supports. This in turns implies that each $\omega$ belongs to both $\Omega_{W_j}$ and $\Omega_{W'_k}$, for some $j$ and $k$. As we have seen, this would imply
\begin{equation}\label{prop}
\sigma(\omega)\propto W_j \propto W'_k.
\end{equation}

However, one can find many examples where no process in one decomposition is proportional to any process in the other. This implies a contradiction and shows that a decomposition such as \eqref{quantont} cannot exist for all CP maps and quantum processes. As a particular example to show the above contradiction, consider a process $W$ corresponding to a quantum channel from a region with a two-level output, $A_O$ to a region with a two-level input, $B_I$:
\begin{equation}
W =  \sum_j q_j W_j = \sum_k p_kW'_k,
\end{equation}
formed from the following two combinations of extremal processes:
\begin{align}
W_1=&\Proj{\id} = \id +X\otimes X -Y\otimes Y+Z\otimes Z,\\
W_2=&\Proj{X}= \id +X\otimes X +Y\otimes Y-Z\otimes Z,\\
W_3=&\Proj{Y} = \id -X\otimes X -Y\otimes Y-Z\otimes Z,\\
W_4=&\Proj{Z} = \id -X\otimes X +Y\otimes Y+Z\otimes Z.
\end{align}

\begin{align}
W'_1=&\Proj{U\id} = \id +X\otimes UXU^\dagger -Y\otimes UYU^\dagger+Z\otimes UZU^\dagger,\\
W'_2=&\Proj{UX}= \id +X\otimes UXU^\dagger +Y\otimes UYU^\dagger - Z\otimes UZU^\dagger,\\
W'_3=&\Proj{UY} = \id -X\otimes UXU^\dagger -Y\otimes UYU^\dagger-Z\otimes UZU^\dagger,\\
W'_4=&\Proj{UZ} = \id -X\otimes UXU^\dagger+Y\otimes UYU^\dagger+Z\otimes UZU^\dagger.
\end{align}
where $X, Y$ and $Z$ are the Pauli matrices, $U$ is a unitary, and we used the notation $\Proj{V}:=\sum_{rs}\ket{r}\bra{s}\otimes V \ket{r}\bra{s}V^{\dag}$ for the Choi representation of a unitary $V$.

It is clear that no $W_j $ is proportional to any $W'_k$ for an appropriate choice of $U$, and we have a contradiction with~\eqref{prop}.

\section{Process-contextual extensions of quantum theory}\label{extension}

Contextuality proofs do not always require both preparation and measurement non-contextuality. Indeed, many no-go theorems focus on the requirement of measurement non-contextuality alone. Interestingly, even without preparation non-contextuality, measurement non-contextuality imposes strong constraints on the ontology. Essentially, any non-contextual ontology must reduce to the Beltrametti-Bugajski (BB) model~\cite{beltrametti95}, which identifies elements of reality with the quantum wave function. An important consequence of this result is that no measurement non-contextual extension of quantum theory exists that can provide more accurate predictions of experimental outcomes~\cite{Montina2011}.

It is thus interesting to consider dropping the requirement of process non-contextuality in our framework, leaving instrument non-contextuality as the sole requirement.  It is easy to see that instrument non-contextual, process-\emph{contextual} models are possible. An example is a model where the ontic process is directly identified with the quantum process:
\begin{equation}
g_W\left(\omega\right) =\delta\left(W-\omega\right).
\label{BB}
\end{equation}
Operational probabilities are then recovered simply by using the ``quantum process rule'', Eq.~\eqref{born}, for the ontic frame function:
\begin{equation}
f_{\omega}(M^A, M^B\dots) = \tr \left[\mathbf{M}\, \omega\right].
\end{equation}
This ``crude'' ontological model is similar to the BB model. A difference is that the BB model only identifies \emph{pure} quantum states with elements of reality, while in Eq.~\eqref{BB} \emph{any} process counts as ontic, including those corresponding to mixed states or noisy channels. One could refine the above model by only allowing an appropriately defined ``pure process'' to be ontic. (See however Ref.~\cite{Araujo2017purification} for possible ambiguities regarding such a definition.)

A similar non-extendability result to that of~\cite{Montina2011} also holds in our case. As already discussed above, the only instrument non-contextual frame function must be given by Eq.~\eqref{framenoncontextual}, namely to every ontic process $\omega$ is associated a process matrix $\sigma\left(\omega\right)$. The implication is that an instrument non-contextual hidden variable cannot provide more information than that contained in a process matrix. We thus conclude that quantum mechanics admits no non-trivial, instrument non-contextual extension. Indeed, this result holds independently of any assumptions one may make about the causal structure of a possible underlying ontology. Therefore, even instrument non-contextuality alone poses strong restrictions on hidden variable models that attempt to leverage exotic causal structures to recover a non-contextual notion of reality.

\section{Discussion}

We have shown that it is not possible to construct an ontological model that is both instrument and process non-contextual and also accords with the predictions of quantum mechanics. We take both forms of non-contextuality to be very reasonable assumptions if one wishes some aspect of "reality" to be describable in a manner that is independent of the act of experimentation. Thus our work shows that models that posit unusual causal, global or dynamical relations will not solve a key quantum mystery, that of contextuality. 

Standard no-go theorems show that quantum theory is not consistent with ontological models where the properties of a system exist prior to and independently of the way they are measured. A possible interpretation is that properties \emph{do} exist, but they are in fact dependent on future actions. Here we have shown that hidden variable models that attempt to leverage such influence from the future have to violate some broader form of non-contextuality. This new notion of non-contextuality refers to the rules that dictate how local actions influence observed events, rather than to states and measurements.

We have introduced three assumptions in order to analyse non-contextuality in such scenarios where influence from the future is possible. The core idea is captured by the assumption of $\omega$-mediation. This states that an agent's actions should effect the world according to rules or laws that do not themselves depend on such actions. 
Indeed, if the rules changed every time we changed how we intervened on the world, we would not call them ``rules'' to begin with. In the context of ontological models, this assumption allows one to assume that experiments uncover an aspect of nature that is unchanging.  

The second assumption, instrument non-contextuality, states that operationally equivalent interventions should not produce distinct effects at the ontological level. We have shown that this assumption is compatible with scenarios that \emph{would} be interpreted as contextual when viewed from an ordinary, time-oriented perspective. For example, we have illustrated that time-travelling models where states \emph{can} depend on future interventions satisfy the requirement of instrument non-contextuality. Despite this generality, instrument non-contextuality is nonetheless sufficient to rule out all non-trivial hidden-variable extensions of quantum theory: 
Any additional variable that could provide better predictions for quantum statistics than ordinary quantum mechanics must be instrument \emph{contextual}.

Our third assumption, process non-contextuality, states that the probabilistic assignment of the ontic description of an experiment should reflect the operationally equivalent arrangements of the same experiment. Here by ``experiment'' we mean the specification of the set of conditions under which agents can operate. That is, we include in this description all aspects of a physical scenario other than the choices of settings and the observed outcomes. Such aspects include what kind of systems are involved, the laws describing such systems, boundary conditions, etc. We have shown that no ontic model can satisfy this requirement of process non-contextuality, including those that directly identify quantum objects as ontic. 

The distinction between background environment variables and locally controllable settings  that one makes when describing experiments using our approach is of course mobile. What counts as a freely chosen parameter in one situation can count as a fixed parameter in another. Our result is robust under such a shift in perspective: no matter how we decide to describe a quantum experiment, it will not be possible to find an ontic representation for it that is both instrument and process non-contextual.

Finally, we draw attention to the fact that our results rely on complete matching to the operational predictions of quantum theory. This is a recognised feature of all ontological models that rely on operational equivalence classes and leaves open the possibility that particular ontological models might allow for some experimentally testable, different predictions. Thus, for proponents of particular retrocausal models, the door remains open to develop their ontology such that they can predict some possible deviation from quantum statistics. In the face of such statistical deviation, the possibility of a non-contextual ontological model remains open.


\vspace{-2pt}
\begin{acknowledgments}
\vspace{-2pt}

We thank {\v C}aslav Brukner, Eric Cavalcanti, Ravi Kunjwal, Matthew Leifer, Gerard Milburn, Alberto Montina, David Schmid, Robert Spekkens, and Ken Wharton for helpful discussions. This work was supported by an Australian Research Council Centre of Excellence for Quantum Engineered Systems grant (CE 110001013), and by the Templeton World Charity Foundation (TWCF 0064/AB38). F.C.\ acknowledges support through an Australian Research Council Discovery Early Career Researcher Award (DE170100712). This publication  was made possible through the support of a grant from the John Templeton Foundation. The opinions expressed in this publication are those of the authors and do not necessarily reflect the views of the John Templeton Foundation. We acknowledge the traditional owners of the land on which the University of Queensland is situated, the Turrbal and Jagera people.
\end{acknowledgments}



\providecommand{\href}[2]{#2}
\raggedright

\appendix

\section{Wharton's retrocausal toy model}
\label{toywharton}

The core idea of the model is to represent a system across space-time, analogously to the representation of a system in space in thermodynamical equilibrium. Rather than being determined by dynamical evolution, the states at each point in space-time are known with some probability. This is similar to how macrostates can be considered as providing probability distributions for microstates. 

In this model each event in space-time is represented as a site, labelled by the index $j$, within a lattice. At each site $j$  we can have a particle in a state $\lambda$, whose possible values are assumed to be $\pm 1$ for simplicity. The entire system across space-time is treated ``all-at-once'' in the same way one would treat a spatially extended system, where each site represents a different location in space. The system is then associated with a Hamiltonian $H=-\sum_{<i,j>} \lambda_i \lambda_j$, where the sum is taken over nearest-neighbours according the geometry of the lattice. All we know about the system is that it is in a thermal state, with inverse temperature $\beta$, thus the probability for a certain configuration $\vec{\lambda}:=\left(\lambda_1,\lambda_2,\dots\right)$ is $p(\vec{\lambda}|\beta) \propto e^{- H(\vec{\lambda})\beta}$. If we learn the state of one of the sites, we need to update the thermal distribution by conditioning on the observed value. However, since the model is supposed to represent a space-time configuration, the sites we can observe at any given time are restricted. 
\begin{figure}[ht]%
\includegraphics[width=0.8\columnwidth]{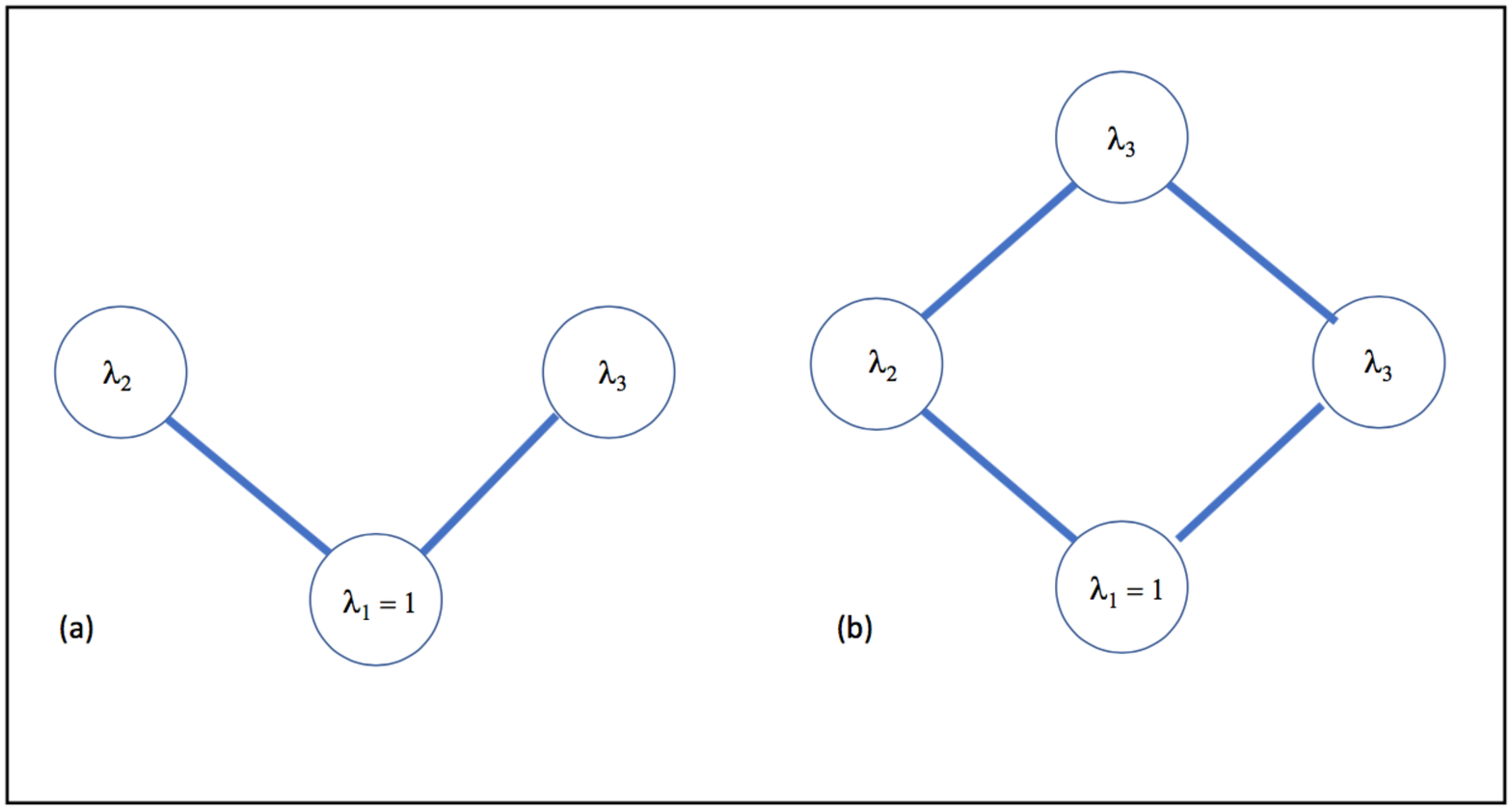}%
\caption{\textbf{Wharton's toy model}~\cite{Wharton2014}. Each node $j$ represents a location in space-time where a system can be found in a state $\lambda_j$, $j=1,2,\dots$. The state of the entire system is sampled from a thermal ensemble, defined by a Hamiltonian containing interactions between nodes connected by an edge, where each node is treated as a site in a spatially distributed lattice. (a) Observing the system at a given time reveals the state at one of the nodes, e.g.\ $\lambda_1=1$, upon which the probability assignment at the other nodes has to be updated. (b) The analogue of an interference experiment is represented by the insertion of an additional node in the future, which results in a different thermal state and thus in a different probability distribution for all states. An observer at an earlier time that ignores this possibility might interpret such a dependence from future actions as a form of contextuality.}%
\label{spins}%
\end{figure}

Retrocausality is introduced by assuming that performing a measurement at any given time can result in the introduction of a new site, thus changing the geometry of the system, Fig.~\ref{spins}. Assuming a thermal state with a given temperature, the two geometries result in different probability distributions for the microstates. If the system is interpreted as time oriented, and the influence of the future intervention is ignored, then one might be led to the conclusion that it is impossible to assign non-contextual states of reality to the system. The analogy is seen with a quantum interference experiment, where a measurement in the future is assumed to change the conditions that determine the state of the system in its past. If the influence from the future measurement is included, argues Wharton, then one might be able to recover an ontic interpretation of quantum mechanics, where the quantum state simply represents lack of information about the underlying state.

This model is interesting because causal influence is not mediated by an explicit mechanism, as opposed to ordinary dynamical systems including the time-travelling examples in the main text. Nonetheless, it is possible to fit this model into our general ontological framework, where the observed probabilities are mediated by an ontic process. Crucially, the model turns out to be both instrument and process non-contextual, showing that approaches of this type cannot reproduce the predictions of quantum theory.

\subsection*{Classical systems on an arbitrary geometry}
We consider a more general version of Wharton's model, with arbitrary geometry, an arbitrary set of discrete values for the states, and arbitrary local interactions.

Consider a set $\mathcal{N}$ of $\left|\mathcal{N}\right|=N$ sites. Each site $j\in\mathcal{N}$ can contain a classical system whose state $\lambda_j$ can take value in some set $\mathcal{S}_j$. 
The state of the entire system is thus described by a vector $\vec{\lambda}\equiv\left(\lambda_1,\dots,\lambda_N\right)\in \mathcal{S}:=\bigtimes_{j\in\mathcal{N}}\mathcal{S}_j$.

A Hamiltonian function $H(\vec{\lambda})$ is defined on the system.
We assume that this Hamiltonian is \emph{local}, namely it is a sum of terms representing local interactions between sites. A subset of sites $e\subset \mathcal{N}$ contributing to an interaction term is called ``hyperedge'' and the set $\mathcal{E}$ of hyperedges defines a ``hypergraph'' over $\mathcal{N}$. The Hamiltonian can thus be decomposed as
\begin{equation}
H=\sum_{e\in\mathcal{E}} h_e,
\label{local}
\end{equation}
where each term $h_e$ is function on the space $\mathcal{L}_e:=\bigtimes_{j\in e}\mathcal{S}_j$. By convention, we identify the state $\lambda_j=0$ of system $j$ with the ``empty site'', namely with no system in it. This implies that, for every hyperedge $e$ containing $j$, 
\begin{equation}
h_e\left(\dots,\lambda_j=0,\dots\right)=0.
\label{void}
\end{equation}
In other words, each interaction term vanishes when one of the sites on which it acts is empty. In this way, ``different geometries'' corresponding to additional or missing sites, are simply represented as a particular choice of states in a fixed geometry.
 
In our terminology, each site $j$ represents a (space-time) region and each state $\lambda_j$ represents an event. We can interpret each event as ``ontic''; however, since we assume that each ontic event can also be observed, ontic and operational events are identified.

\paragraph*{No control.}
Before considering the possibility of interventions, it is useful to see how our framework applies to the simpler scenario with no interventions. In this case, ``process'' is synonymous with ``state''. Thus, a deterministic process is simply a specific microstate $\vec{\lambda}$, while a general probabilistic process is a probability distributions $P\left(\vec{\lambda}\right)$. For the case of a thermal state, where the only information we can access about the environment is the inverse temperature $\beta$, the operational process is probabilistic, given by the Gibbs distribution
\begin{equation}
p(\vec{\lambda}\mid \beta) = \frac{e^{-\beta H\left(\vec{\lambda}\right)}}{Z(\beta)},\quad  Z(\beta) = \!\!\sum_{\vec{\lambda}\in\mathcal{S}} e^{-\beta H\left(\vec{\lambda}\right)}.
\label{Gibbs}
\end{equation}

Since there are no irrelevant environment variables in this model, questions about contextuality do not arise: each value of the environment variable corresponds to just one process (i.e.\ to one probability distribution for the ``events''). Therefore, at the formal level, we could identify the operational process with the ontic process; the resulting model would be process non-contextual by construction (instrument non-contextuality is even more trivial here, because there is no choice of instruments). A more natural ontological model is a deterministic one, where each ontic process (or ontic state) is identified with one microstate $\vec{\lambda}$. As required by the general formalism, the operational process provides a probability distribution over the possible ontic processes, and knowing the ontic process makes knowledge of the operational process redundant (the ontic process ``screens off'' the operational one), in agreement with the property of $\omega$-mediation.

\paragraph*{Local control.}
Local instruments are defined as subsets of events and represent the possibility of local control. Thus, in general, the possible sets of instruments at a site $j\in\mathcal{N}$ corresponds to a subset $\mathfrak{I}^j\subset \mathcal{S}_j$.
As a simple case-study, we consider the scenario where the only control is inserting or removing a site, as in Wharton's example. Therefore, for each region $j\in\mathcal{N}$ there are two possible instruments:
\begin{equation}
\mathfrak{I}^j_0:=\left\{0\right\},\qquad \mathfrak{I}^j_1:=\mathcal{S}_{j}{\setminus\left\{0\right\}}.
\label{instruments}
\end{equation}
A prominent feature of this example is that instruments are disjoint sets, so there is never an event that belongs to two distinct instruments. This ensures the instrument non-contextuality of the model.

As in the no-control case, a deterministic process corresponds to a specification of all events, while a probabilistic process corresponds to a probability distribution for the possible events. The possibility of control means that the events now can depend on the instruments, so the process must encode this dependency. Thus, a deterministic process is given by a set of functions
\begin{equation}
\vec{\mathfrak{I}}\equiv\left(\mathfrak{I}^1,\dots,\mathfrak{I}^N\right)\mapsto \vec{\lambda},\quad \lambda_j=\omega_j(\vec{\mathfrak{I}}),\,j\in\mathcal{N}
\label{detprocess}
\end{equation}
such that, for each $j\in\mathcal{N}$, 
\begin{equation}
\omega_j\left(\dots, \mathfrak{I}^j=\mathfrak{I}^j_0,\dots\right)=0,\qquad \omega_j\left(\dots, \mathfrak{I}^j=\mathfrak{I}^j_1,\dots\right)\neq 0.
\label{consistency}
\end{equation}
Condition \eqref{consistency} simply says that if we choose to remove the system from site $j$ ($\mathfrak{I}^j=\mathfrak{I}^j_0$), then there will be no system at site $j$ ($\lambda_j=0$), while if we choose to insert the system ($\mathfrak{I}^j=\mathfrak{I}^j_1$) then the system will be there, in one of its possible states ($\lambda_j\in \mathcal{S}_{j}{\setminus\left\{0\right\}}$).

For a probabilistic process, dependency on the instruments is encoded in a conditional probability distribution $p(\vec{\lambda}\mid \vec{\mathfrak{I}})$. The probabilistic version of the consistency condition \eqref{consistency} reads
\begin{align}\label{probconsistency}
P&\left(\dots, \lambda_j\neq 0,\ldots \mid \dots, \mathfrak{I}^j=\mathfrak{I}^j_0,\dots\right)=0,\\ \nonumber
P&\left(\dots, \lambda_j = 0,\ldots \mid \dots, \mathfrak{I}^j=\mathfrak{I}^j_1,\dots\right)=0.
\end{align}

A more compact way to represent a (non-contextual) process is through a frame function, which can be defined piece-wise as:
\begin{equation}
f(\vec{\lambda}) := 
p(\vec{\lambda}\mid \vec{\mathfrak{I}}) \quad \textrm{for } \lambda_1\in\mathfrak{I}^1, \dots, \lambda_N\in\mathfrak{I}^N.
\label{frame}
\end{equation}
The consistency condition \eqref{probconsistency} is then expressed as
\begin{equation}
p(\vec{\lambda}\mid \vec{\mathfrak{I}}) = f(\vec{\lambda}) \prod_{j\in \mathcal{N}}\chi_{\mathfrak{I}^j}(\lambda_j).
\label{frameconsistency}
\end{equation}

Let us pause for a moment on this definition. The frame function is defined as a function $f:\mathcal{S}\rightarrow [0,1]$. That is, it assigns a probability to each $N$-tuplet $\vec{\lambda}=\left(\lambda_1\dots,\lambda_N\right)$, without needing any additional information about the instruments. In the case we are considering, different instruments are non-overlapping sets of states. Therefore, if we know the state $\lambda_j$, we automatically know the instrument $\mathfrak{I}^j$ and requiring ``independence from the instruments'' is completely trivial: either the site is there, and the instrument is $\mathfrak{I}^j_1$, or the site is not there, and $\mathfrak{I}^j = \mathfrak{I}^j_1$. Once we know the state, there is nothing more the instrument can tell us. Technically, each value $\vec{\mathfrak{I}}$ defines a subset in the domain of $f$, and the value $f$ takes in each of these subsets is given by the conditional probability \eqref{frame}. The non-overlapping of different instruments is crucial for this construction: if the same $\vec{\lambda}$ could belong to two different instruments, we would not know which value of $p(\vec{\lambda}\mid \vec{\mathfrak{I}})$ to use to define the frame function. For overlapping instruments, the existence of a frame function is equivalent to the assumption of instrument non-contextuality.

\paragraph*{Processes for the thermal state}
Once again, the only environment variable is the inverse temperature $\beta$, which thus parametrises the operational processes. Given the above discussion, it should be clear that, for each $\beta$, we can write a conditional probability in the form \eqref{frameconsistency}, where the frame function is defined as in Eq.~\eqref{frame}, with probabilities provided by the Gibbs distribution \eqref{Gibbs}. Explicitly,
\begin{align} \label{framethermal}
f_{\beta}(\vec{\lambda}) =& 
\frac{e^{-\beta H(\vec{\lambda})}}{Z(\beta\mid \vec{\mathfrak{I}})}\quad \textrm{ for } \vec{\lambda}\in \vec{\mathfrak{I}},\\ \textrm{\hspace{-60pt} where }
Z(\beta\mid \vec{\mathfrak{I}}):=&\!\!\sum_{\vec{\lambda}\in\vec{\mathfrak{I}}} e^{-\beta H\left(\vec{\lambda}\right)}.
\end{align}

Let us stress that, from the perspective of our framework, we might as well stop here: we already have a model that is both instrument and process non-contextual. The point of our theorem is to see if it is possible to write a given operational model in terms of an underlying non-contextual model; if that is possible, the operational model cannot reproduce the predictions of quantum mechanics. In this case we already have a non-contextual model, so we know it cannot reproduce quantum mechanics. Note that the theorem does not rely on any interpretation we might assign to ontic processes, events etc; it is simply a statement about properties that an ontological model can or cannot have. 

For the sake of completeness, and since we would more naturally associate ontology with determinism,  we can write explicitly how a deterministic process model looks in the present case study.
Recall that a deterministic process is a (multi-valued) function $\vec{\omega}\equiv \left(\omega_1\dots \omega_N\right)$  from the instruments to the events. For notational convenience, we can identify the two possible instruments $\mathfrak{I}^j_{x_j}$ at each site $j$ with their label $x_j\in\left\{0,1\right\}$. Therefore, a choice of instruments is given by $N$ binary variables $\vec{x}\equiv\left(x_1,\dots,x_N\right)$ and a process is identified with $2^N$ $N$-tuples $\left\{\vec{a}_{\vec{x}}\right\}_{\vec{x}\in \left\{0,1\right\}^N}$, where $\vec{a}_{\vec{x}}:= \vec{\omega}(\vec{\mathfrak{I}}_{\vec{x}})$. 
A deterministic ontological model is thus defined by a conditional probability distribution 
\begin{equation}
p\left(\vec{\omega}\mid\beta\right)\equiv P\left(\left\{\vec{a}_{\vec{x}}\right\}_{\vec{x}\in \left\{0,1\right\}^N} \mid\beta\right)
\end{equation} 
that reproduces the operational probabilities via $\omega$-mediation:
\begin{equation}
\sum_{\left\{\vec{a}_{\vec{x}}\right\}_{\vec{x}\in \left\{0,1\right\}^N} }
p\left(\vec{\lambda}\mid \vec{\mathfrak{I}}, \left\{\vec{a}_{\vec{x}}\right\}_{\vec{x}\in \left\{0,1\right\}^N}\right) p(\left\{\vec{a}_{\vec{x}}\right\}_{\vec{x}\in \left\{0,1\right\}^N} \mid \beta) 
= p\left(\vec{\lambda}\mid \vec{\mathfrak{I}}, \beta\right),
\label{omegasep}
\end{equation}
%
where the sum should be understood as
\begin{equation}
\sum_{\left\{\vec{a}_{\vec{x}}\right\}_{\vec{x}\in \left\{0,1\right\}^N} }\equiv \sum_{a^1_{0} \in \mathfrak{I}^1_{0}} \sum_{a^1_{1} \in \mathfrak{I}^1_{1}} \dots \sum_{a^N_{0} \in \mathfrak{I}^N_{0}} \sum_{a^N_{1} \in \mathfrak{I}^N_{1}}
\end{equation}
and the ``ontic'' probabilities are given by
\begin{align} \label{detprob}
P\left(\vec{\lambda}\mid \vec{\mathfrak{I}}, \left\{\vec{a}_{\vec{x}}\right\}_{\vec{x}\in \left\{0,1\right\}^N}\right) 
&= \prod_{j\in \mathcal{N}}\chi_{\mathfrak{I}^j}(\lambda_j) 
\sum_{\vec{x}\in \left\{0,1\right\}^N} \delta_{\vec{\lambda}\,\vec{a}_{\vec{x}}}, \\
\delta_{\vec{\lambda}\,\vec{a}_{\vec{x}}} &
:= \prod_{j\in \mathcal{N}} \delta_{\lambda_j\,(a_{\vec{x}})_j}.
\end{align}
%

We now show that the conditional probabilities for the ontic process in our thermal model are given by
\begin{equation}
P\left(\left\{\vec{a}_{\vec{x}}\right\}_{\vec{x}\in \left\{0,1\right\}^N} \mid\beta\right) = \prod_{\vec{x}\in\left\{0,1\right\}^N}f_{\beta}(\vec{a}_{\vec{x}}),
\label{onticprob}
\end{equation}
where the operational frame function $f_{\beta}$ is given by expression \eqref{framethermal}. To see that the conditional probabilities \eqref{onticprob} provide an ontological model for the original thermal-state model, one can verify that, by putting together expressions \eqref{onticprob} and \eqref{detprob} into Eq.~\eqref{omegasep}, one indeed obtains the operational probabilities~\eqref{frameconsistency}.
 Explicitly, 
\begin{align*}
\sum_{\left\{\vec{a}_{\vec{x}}\right\}_{\vec{x}\in \left\{0,1\right\}^N}}&
p\left(\vec{\lambda}\mid \vec{\mathfrak{I}}, \left\{\vec{a}_{\vec{x}}\right\}_{\vec{x}\in \left\{0,1\right\}^N}\right) p(\left\{\vec{a}_{\vec{x}}\right\}_{\vec{x}\in \left\{0,1\right\}^N} \mid \beta) \\
=&
\prod_{j\in \mathcal{N}}\chi_{\mathfrak{I}^j}(\lambda_j) 
\sum_{\left\{\vec{a}_{\vec{x}}\right\}_{\vec{x}\in \left\{0,1\right\}^N}}
\sum_{\vec{x}'\in \left\{0,1\right\}^N} \delta_{\vec{\lambda}\,\vec{a}_{\vec{x}'}} \prod_{\vec{x}\in\left\{0,1\right\}^N}f_{\beta}(\vec{a}_{\vec{x}}) \\
=& \prod_{j\in \mathcal{N}}\chi_{\mathfrak{I}^j}(\lambda_j) 
\sum_{\vec{x}'\in \left\{0,1\right\}^N} 
\sum_{\vec{a}_{\vec{x}'}\in\vec{\mathfrak{I}}_{\vec{x}'}}
\delta_{\vec{\lambda}\,\vec{a}_{\vec{x}'}} 
\left[\sum_{\left\{\vec{a}_{\vec{x}}\right\}_{\vec{x}\neq \vec{x}'}}\prod_{\vec{x}\in\left\{0,1\right\}^N}f_{\beta}(\vec{a}_{\vec{x}})
\right]\\
=& \prod_{j\in \mathcal{N}}\chi_{\mathfrak{I}^j}(\lambda_j) 
\sum_{\vec{x}'\in \left\{0,1\right\}^N} 
\sum_{\vec{a}_{\vec{x}'}\in\vec{\mathfrak{I}}_{\vec{x}'}}
\delta_{\vec{\lambda}\,\vec{a}_{\vec{x}'}} f_{\beta}(\vec{a}_{\vec{x}'})
\left[\prod_{\vec{x} \neq \vec{x}'} \sum_{\vec{a}_{\vec{x}}\in\vec{\mathfrak{I}}_{\vec{x}}}
f_{\beta}(\vec{a}_{\vec{x}})
\right] \\
=& \prod_{j\in \mathcal{N}}\chi_{\mathfrak{I}^j}(\lambda_j) 
 f_{\beta}(\vec{\lambda}), 
\end{align*}
where we used the normalisation of the frame function, $\sum_{\vec{\lambda}\in\vec{\mathfrak{I}}_{\vec{x}}}
f_{\beta}(\vec{\lambda})=1$ for every collection of instruments $\vec{\mathfrak{I}}_{\vec{x}}\equiv \left(\mathfrak{I}^1_{x_1},\dots,\mathfrak{I}^N_{x_N}\right)$.

\end{document}